# Effect of sonication time and surfactant concentration on improving the bio-accessibility of lycopene synthesized in poly-lactic co-glycolic acid nanoparticles

Anwar Ul Alam, M.[1]; Kassu, A.[2]; and Kassama, K.[3]*

* Corresponding Author

[1]Department of Food Science
Cornell University

[2]Department of Mechanical, Civil Engineering & Construction Management,
Alabama A & M University, Meridian St, Normal, Huntsville, AL-35762

[3]Department of Food and Animal Sciences,
Alabama A & M University, Meridian St, Normal, Huntsville, AL-35762

**Highlights**

- PLGA based nanoencapsulation prevent lycopene degradation
- Physicochemical properties determine bioaccessibility of lycopene-nanoparticles
- Control release kinetics of lycopene followed Non-Fickian type of diffusion
- PLGA improve lycopene bioaccessibility by 3-9 times.

**ABSTRACT**. The use of biodegradable polymers simplifies the development of therapeutic devices with regards to transient implants and three-dimensional platform suitable for tissue engineering. Further advances have also occurred in the controlled released mechanism of bioactive compounds encapsulated in biodegradable polymers. This application requires the understanding of the physicochemical properties of the polymeric materials and their inherent impact on the delivery of encapsulated bioactive compounds. Researcher evidenced that sonication time and surfactant concentration have potential effect on physicochemical properties of encapsulated nanoparticles. Hence, the objective of this study was to evaluate the effect of surfactant and sonication time on the bio-accessibility of lycopene encapsulated in polymeric nanoparticles. The emulsion evaporation method was used to encapsulate lycopene in poly-lactic co-glycolic acid (PLGA). Optimization of physicochemical properties of encapsulated nanoparticles has done by modifying surfactant concentration, sonication time and polymer concentration during the emulsion evaporation technique. Physicochemical and morphological characteristics were measured with a zetasizer and SEM (Scanning Electron Microscopy) technique, while the encapsulation efficiency and controlled release kinetics was determined by spectrophotometric, and bioaccessibility by dialysis method. The results have shown that sonication time had significant ($p < 0.05$) influenced on encapsulation efficiency, hydrodynamic diameter, and stability of nanoparticles. However, the sonication time strongly controls the stability (zeta potential value) and encapsulation efficiency (63.8% and 85.7%, respectively) of the lycopene nanoparticles where surfactant (-20.3%) and polymer (49.9%) concentration strongly ($P<0.05$) control the hydrodynamic diameter and conductivity of the nanoparticles. Sonication time of 5 minutes, surfactant concentration of 300 mg and polymer concentration of 300 mg observed to be the optimum condition for physicochemical properties suitable for better bioaccessibility of nanoparticles. During the 12 days, optimized nanoparticles released 53.89 ± 6.65 % of the total content which was fairly constant rate of 4.49 percent/day into the invitro phosphate buffer solution (PBS, pH =7.4) system at 37°C. Burst release was observed on the 13th day, where 41.7 ± 3.75 % of the total content was released within a single day. Optimized nanoparticles had a bioaccessibility of 90.87±3.99% which is 3-9 times greater compared to non-encapsulated lycopene ingested through the oral route. Furthermore, both control release kinetics (R=96%) and absorption (R=97.46%) profiles of the lycopene nanoparticles followed non-Fickian types of diffusion. This study will have significant impact on the manufacturing of functional food with encapsulated ingredients and provide an understanding of their inherent absorption and control release mechanism in the GIT and blood stream for subsequent targeted delivery of lycopene.

# 1. INTRODUCTION

The development of biodegradable polymeric materials in biomedical applications has markedly advanced during the last half-century. Moreover, the polymeric biomaterials sector is expected to show a 22.1% growth rate in the Global implantable biomaterial market due to its promising potential in biomedical applications (Taylor et al., 2015). Biodegradable polymeric materials mostly facilitate the development of therapeutic devices concerning transient implants and three-dimensional platforms for tissue engineering. Further advancements have occurred in the controlled delivery of bioactive compounds like lycopene for preventing the chronic diseases from the human body. These applications require analyzing the polymeric materials' physicochemical, biological, and degradation properties, which can help identify the therapeutic agent's delivery efficiency to the targeted site. As a result, researchers have designed a wide range of natural or synthetic polymers which can undergo hydrolytic or enzymatic degradation in a controlled way and play a potential role in delivering bioactive compounds for biomedical applications (Song et al., 2018). Polylactic co-glycolic acid (PLGA) is a prominent biodegradable polymer in biomedical applications and can be modified with varying lactide/glycolide ratios to achieve transitional degradation rates between polylactic acid (PLA) and poly-glycolic acid (PGA). Generally, the copolymer PLGA offers more incredible control release properties than its co-constituent polymer by varying its monomer ratio (Felix Lanao et al., 2013). Additionally, Shi et al. (2009) reported that PLGA/hydroxyapatite microsphere composites significantly improve osteoblast proliferation by upregulating an essential osteogenic enzyme named as alkaline phosphatase. Controlled release of the biodegradable polymer in biomaterial delivery can be described by the Higuchi model when biomaterials are water-soluble or low soluble and incorporated in semisolid and solid matrices. On the other hand, first-order model explains the concentration-dependent release kinetics (Higuchi, 1963; Suvakanta et al., 2010). Hixson-Crowell cube root model describes the release kinetics by dissolution rate of molecules (Hixson and Crowell, 1931) where the Gallagher-Corrigan model is a mathematical model that can explain a single fraction of bio-actives released from the biodegradable polymeric capsules (Gallagher et al., 2000; Balcerzak et al., 2010; Arezou, 2015). Korsmeyer et al. (1983), Ritger and Peppas (1987), and Korsmeyer and Peppas (1984) established experimental equation to explain both Fickian and non-Fickian release of bioactive compounds from swelling as well as non-swelling polymeric delivery systems.

Lycopene belongs to the tetraterpene carotenoid family, found in red fruit and vegetables. Eleven conjugated double bonds provide its exceptional potentiality to scavenge lipid peroxyl radicals, reactive oxygen species, and nitric oxide which further contributes chronic diseases prevention. However, the lipophilic fractions observed an increased antioxidant activity, ranging from 9% to 40%, after gastric digestion but a sharp decrease due to lipase enzymes in the intestine (Tommonaro et al., 2017). Similarly, light, heat, and oxygen can influence the degradation of lycopene during the processing of foods to ensure its microbiological safety. However, there is a poor statistical association between dietary intake and serum lycopene levels if ingested orally due to its enzyme sensitivity and poor absorption profile in human GIT (Gastrointestinal Tract). Hence, it is very improbable that nutritional intervention alone could be instrumental in correcting lycopene deficiency or chronic diseases prevention. Therefore, new nutraceutical formulations of lycopene or other carotenoids with enhanced bioavailability are urgently needed to support the prevention of chronic diseases (Petyaev, 2016). Researchers proved that encapsulating in the biodegradable polymer at the nano-level might be an attractive platform to improve the bio-accessibility, control release, and targeted delivery of bioactive compounds without producing any toxicity (Liarou et al., 2018; Nguyen et al., 2018; Ye et al., 2018;

Lee et al., 2013). Emulsion evaporation was the most efficient and reliable method to encapsulate bioactive compounds at the nano-level (Kassama and Mishir, 2017).

Physicochemical properties of the nanoparticles can be a potential indicator for bioaccessibility, delivery efficiency, control release, and targeted delivery of bioactive compounds encapsulated in polymeric nanoparticles (Kassama and Mishir, 2017). Nanoparticles have a size of below 100 nm are highly absorbed through the intestinal mucosa lining (Lai et al., 2007), where between 300 and 500 nm have poor (Norris et al., 1998), and particles of greater than 500 nm have no absorption through the intestinal lining (Norris et al., 1998). No charge and negative charge of nanoparticles have high and low absorption profiles, respectively (Lai et al., 2009). However, a negative charge hinders biliary excretion, whereas positive charge help adheres to mucus membrane and thus increase the absorption (Bertrand & Leroux, 2012). Kassama and Mishir (2017) observed that surfactant, biodegradable polymer concentration, and sonication time significantly affect the physicochemical properties of the nanoparticles. Hence, it is hypothesized that surfactant, polymer concentration, and sonication time will significantly affect the physicochemical properties and bioaccessibility of lycopene synthesized in PLGA nanoparticles. The main objective of this study is to verify the effect of surfactant, polymer concentration, and sonication time on the physicochemical and bioaccessibility of the lycopene nanoparticles.

## MATERIALS AND METHODS
### 2.1 Research design:

In this study, a three-factor factorial experimental design was used to evaluate the sonication time, surfactant, and polymer concentration effect on physicochemical properties of lycopene-PLGA nanoparticles. Surfactant concentration (at 300, 400, and 500 mg), sonication time (at 4, 5, and 6 min), and biodegradable polymer concentration (at 300, 400, and 500 mg) were three experimental variables where physicochemical properties, encapsulation efficiency, bioaccessibility, and control release kinetics were the main response variables (Figure-1). ANOVA and Tukey test were done at a 5% level of significance. Additionally, multiple regression was done to identify the specific factorial effect on the formulation of nanoparticles.

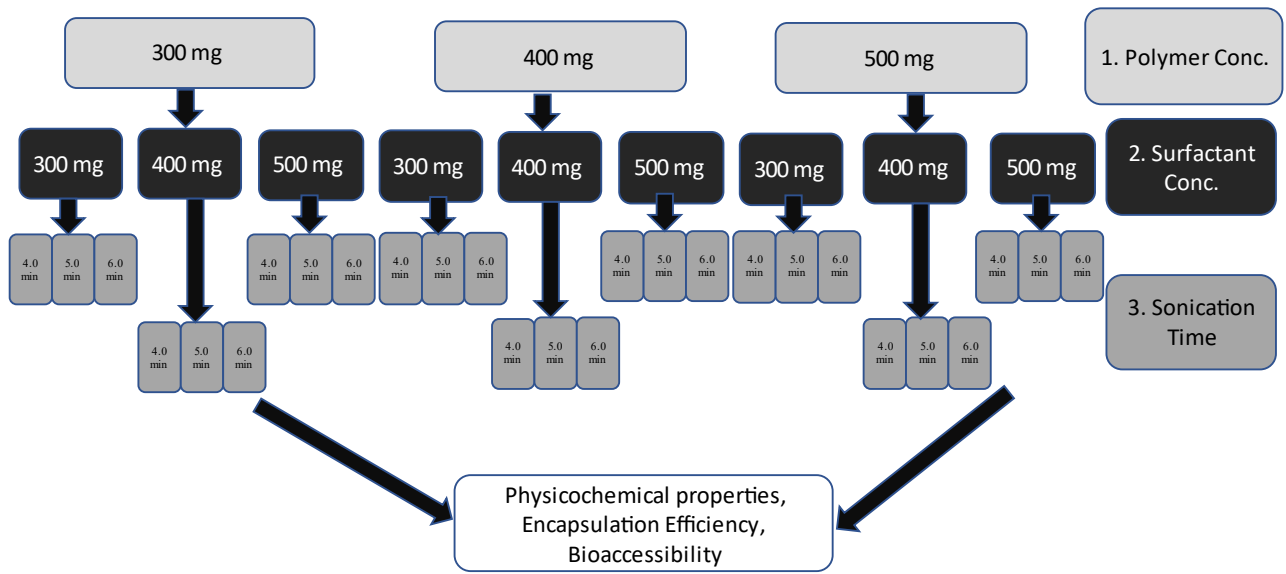
3

Figure 1. Research design for synthesis of lycopene nanoparticles

## 2.2 Extraction of lycopene from tomatoes by column chromatography method

Roma tomatoes were collected from the local markets and cut into small pieces to increase the surface area of the tomatoes. Tomato pieces were placed in a tray of tray dryer and heated overnight at 45°C temperature. Dried tomatoes were ground with solid grinder and took 8 g of ground dried tomato powder for extracting lycopene from it. After dissolving the 8 g of tomato powder with acetone, petroleum, & dichloromethane (1:2:1), filter the tomatoes paste through Whatman filter paper. Tomato pastes were squeezed to release maximum lycopene from the tomato paste. The filtrate then transferred into the preset column. Next, petroleum ether, dichloromethane in 1:1 proportion, and small amount of 95% petroleum ether into the column and kept the solvent mix to settle down into different bands. Red band, distinct for lycopene, was collected into a different flask by evaporating the solvent in a rotary evaporator. After collecting the dry extract, acetone was used to reconstitute the solution to check the purity by UV-spectrophotometric method at 505 nm of wavelength and used as a stock solution for subsequent encapsulation.

## 2.3. Synthesis of lycopene loaded PLGA nanoparticles

The emulsion evaporation method, developed by Kassama and Misir et al. (2017) with some modification, was used to synthesize PLGA-lycopene nanoparticles. Two different phases, organic and water, were made to formulate the encapsulated lycopene-PLGA nanoparticles. PLGA and lycopene solution (40 mg/100ml) were dissolved in ethyl acetate (9 mL) in the organic phase where dimethylamine boren was dissolved in deionized water (60 mL) in the water phase. Water solution was transferred to the organic solution at room temperature and mixed under magnetic stirring. Mixed solution then homogenized by sonication (QSonica 700, QSonica LLC) for 4, 5, and 6 min, respectively at 70% aptitude. After the sonication, homogenized polymeric solutions were run through rotary evaporation (RE 301, Yamato Scientific Co. LTD, Tokyo, Japan) at 20 rpm to evaporate organic solvent (ethyl acetate) from the solution at 35°C temperature (Kassama and Misir, 2017). After evaporation, nanoparticles were collected by subsequent centrifugation and freeze-drying to get solid powder particles. Centrifugation was done at 10,000 rpm for 40 minutes (Farrag et al., 2018) by maintaining the centrifugation temperature in refrigeration range (4°C) to evaluate the encapsulation efficiency where the freeze drying was done to analyze its morphological characteristics.

## 2.4. Measurement of physicochemical properties of nanoparticles

A lab-grade Zeta-sizer (Nano ZS90, Malvern Instrument Ltd, Worcestershire, UK) was used to determine the average hydrodynamic diameter and polydispersity index of lycopene nanoemulsions at 25°C with a scattering angle of 90°. Plastic cuvette of three milliliters was filled up with lycopene nanoemulsion and placed into the sample chamber of the Zeta-sizer to measure the size and polydispersity index of the nanoparticles. On the other hand, zeta potential or the stability, mobility, and electric conductivity of lycopene nanoemulsions were measured by putting the sample into a "U" shape cuvette and running it through the commercial zeta-potential and particle size analyzer at 25°C.

## 2.5. Determination of percentage encapsulation efficiency (EE)

Encapsulation efficiency was measured by Farag et al. (2018) method with some modifications. Two milliliters of loaded nanoparticle solution were subjected to high-speed cooling (4°C) centrifugation at 10,000 rpm for 30 min. Then the supernatant solution was mixed with 2 ml of acetone, followed by a UV-Vis spectrophotometric analysis at 505 nm. Lycopene content was determined by comparing the sample absorption with the standard lycopene curve (Figure 2). Encapsulation efficiency (EE) was calculated by an equation as follows:

$$EE\% = \frac{\text{(Total amount of lycopene added – Non bound lycopene)}}{\text{Total amount of lycopene added}} \times 100 \qquad (1)$$

## 2.6. Measurement of In-vitro Control Release Kinetics of lycopene NPs

Three hundred milligram of freeze-dried loaded lycopene-NPs was placed into a dialysis bag (MWCO 10 kDa) containing phosphate buffer solution (PBS) of 10 mL which was placed in a glass bottle containing 10 mL of phosphate buffer solution (Swackhamer et al., 2019). Dimethylamine boren was used as an emulsifier, whereas Butylated Hydroxy Toluene (BHT) was used both in sample and receiving chamber of phosphate buffer solution to prevent lycopene oxidation (Rao and Murthy, 2000; Swackhamer et al., 2019). A mechanical agitator was used at 120 rpm speed and 37°C temperature to simulate the in-vitro extracellular fluid movement into the bloodstream (Rao and Murthy, 2000). Two milliliter of sample was taken every day from the receiving PBS solution to determine the lycopene content by UV/Visible spectrophotometric method (UV 1800, Shimadzu, Japan) at 505 nm (Jeevitha and Amarnath, 2013).

## 2.7. Measurement of Bio-accessibility of encapsulated lycopene NP

An automatic in-vitro system was developed to simulate the digestion activity of the mouth, stomach, and small intestine of the human metabolic system. Sample was present in the mouth simulator for 1 minute at a pH of 7. Simulated gastric fluid (SGF) was prepared by the method established by Hedren et al., (2002). Then, lycopene nanoparticles transferred in an automatic stirring bioreactor at 37°C for 2 h to simulate gastric digestion. In the first hour of gastric digestion, stomach simulator adjusted the pH at 1.8 to facilitate the protease enzymes activity. After 1 h of digestion in the stomach, the pH of the chamber was adjusted at 5.8 to facilitate the lipase enzyme activity for 1 hour (Drechsler and Bornhorst, 2018). When gastric digestion had been completed, the sample was transferred to the intestinal chamber, where pH was maintained at 6.8 for 2 hours (Tumuhimbise et al., 2009). The samples were collected every 15 minutes in the 1st hour and every 30 minutes for the second hours of gastric digestion. For determining the intestinal effect, samples were collected every hour from the bioreactor simulating intestinal digestion. Bioaccessibility of the lycopene was evaluated after deactivating the enzymes by the salting-out technique.

## 2.8. Measurement of In-vitro Absorption Profile of Lycopene NP

After completion of intestinal enzymatic activities, nanoparticles are usually absorbed through the duodenum lining. Before absorption, the nanoparticles pass through the protective mucus layer to the underlying epithelium for further hydrolysis by brush border enzymes (Bornhorst et al., 2016). For polymeric nanoparticles, absorption usually happens by passive diffusion (80%) through the duodenum and the average pore size of enterocytes present in the duodenum is 200 nm (Borel and Sabliov, 2014). Hence, to simulate the absorption, a membrane filter of 200 nm was placed between the sample collector chamber of simulated intestinal digestion and the receiving chamber for the sample after absorption. Membrane filtration was done in a vacuum condition, and the flow rate has maintained at 0.7-7 bar pressure to simulate the human absorption (Singh and Heldman, 2014).

The absorption of encapsulated lycopene nanoparticles can be described by the Korsmeyer Peppas model, which is explained in the equation-2.

$$\frac{M_t}{M_\alpha} = kt^n \qquad (2)$$

Where,
$Mt/M\alpha$: fraction of drug released at time $t$
$k$: Korsmeyer-Peppas constant
$t$: time
$n$: release exponent
$R^2$ = correlation coefficient

The release mechanism of the lycopene has been determined based on the release exponent ($n$) value, given in table-1 (Ravi and Mandal, 2015).

Table-1: Interpretation of diffusion mechanisms based on release exponential value of the curve

| Release exponent (n) | Drug transport mechanism |
|---|---|
| n < 0.5 | Quasi-Fickian diffusion |
| n = 0.5 | Fickian diffusion |
| 0.5 < n < 1.0 | Non-Fickian transport |
| n = 1.0 | Cass II transport |
| n > 1.0 | Super case II transport |

### 2.9. Measurement of Antioxidant Properties of Encapsulated Lycopene

Antioxidant properties of lycopene nanoemulsions was determined by the DPPH method developed by Moreno, Larrauri & Saura-Calixto, (1998) with slight modification. In brief, a DPPH concentration of 2 mg/100 mL was used as a control. Samples were prepared by dissolving the nanoparticles with dichloromethane, evaporating the solvent by rotary evaporator, separating the released lycopene by centrifugation, filtration by Whatman filter paper & re-dilution with acetone. For making the plat, 200 µL of DPPH and 40 µL of the sample (and blank) were taken to the 96 well microplate reader with triplicate and deionized water was used as blank. The samples were run through the UV-VIS spectrophotometer (UVmini-1240, UV-VIS Spectrophotometer, Shimadzu) at 517 nm, and data were recorded after 90 minutes of contact time.

## RESULTS AND DISCUSSION

### 3.1. Extraction of lycopene by column chromatography

After extracting the lycopene from tomatoes, the purity of the extract was measured by comparing with standard curve of pure lycopene concentration. Figure-1 explained the standard curve for pure lycopene when it dissolved with 50 mL of distilled water in presence of dimethyl amin boren. The purity of the lycopene extracted by column chromatography was 85%.

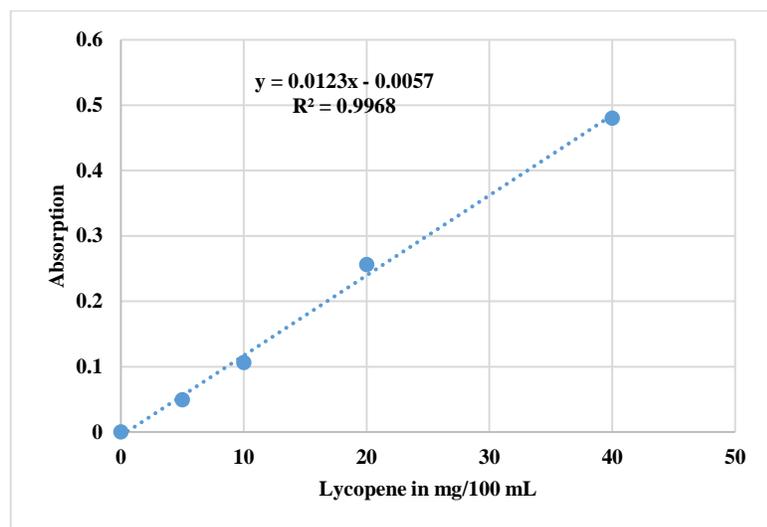

Figure-2. Standard curve for pure lycopene

Standard curve of lycopene had a $R^2$ value of 0.9968. The relation between absorption and lycopene concentration explained by the equation y = 0.0123x -0.0057 where x = concentration of lycopene and y = absorption of lycopene in acetone at 505 nm of wavelength in a UV-Spectrophotometric method.

### 3.2. Factor optimization for nanoparticles formulation

ANOVA table explained that interaction effect of all the independent variables contributes significant changes to the physicochemical properties of formulated nanoparticles.

Table-2: Homogeneity test results for identifying the effect of sonication time, surfactant, and polymer concentration on physicochemical properties of lycopene nanoparticles

| Independent Variables | Physicochemical Properties of Lycopene Nanoparticles | | | | | |
|---|---|---|---|---|---|---|
| | HD (nm) | PDI | ZP (mV) | Mobility (µm cm/cv) | CD (mS/cm) | EE (%) |
| Sonication Time (min) | | | | | | |
| 4 | 201.43[a] | 0.38[a] | -75.57[a] | -5.92[a] | 0.71[a] | 78.06[a] |
| 5 | 204.28[a] | 0.29[c] | -71.88[b] | -5.63[b] | 0.86[b] | 83.47[b] |
| 6 | 175.62[b] | 0.30[ac] | -64.45[c] | -5.05[c] | 0.91[c] | 97.29[c] |
| Surfactant Concentration (mg) | | | | | | |
| 300 | 208.58[a] | 0.29[a] | -72.52[a] | -5.45[a] | 0.80[a] | 87.96[a] |
| 400 | 201.36[b] | 0.33[a] | -69.82[b] | -5.67[b] | 0.81[a] | 87.43[a] |
| 500 | 171.38[c] | 0.35[a] | -69.57[b] | -5.47[a] | 0.87[b] | 83.41[b] |
| Polymer Concentration (mg) | | | | | | |
| 300 | 185.81[a] | 0.32[a] | -71.63[a] | -5.59[a] | 0.67[a] | 85.43[a] |
| 400 | 192.29[b] | 0.33[a] | -71.40[a] | -5.62[a] | 0.78[b] | 85.16[a] |
| 500 | 203.21[c] | 0.33[a] | -68.88[b] | -5.40[b] | 1.02[c] | 88.23[b] |

Footnotes: Different superscript letter within the column means significant differences. All the value is a mean value of triplicates.

Tukey HSD and homogeneity test depicted that sonication time did not affect the polydispersity index of the nanoparticles, where increasing sonication time decreases stability and mobility but improved encapsulation efficiency and conductivity of the nanoparticles significantly (P<0.05). So, for getting high encapsulation efficiency for lycopene, optimized sonication time should be 6 minutes based on the present experimental results. Sonication time had a different effect on the hydrodynamic diameter of the nanoparticles. Increasing sonication time from 4 to 5 minutes did not affect hydrodynamic diameter but had a significant inverse effect at 6 minutes of sonication time. Based on the experimental results, it can be stated that increasing sonication time helped to decrease hydrodynamic diameter which further help the bioavailability as size less than 100 nm has the highest absorption profile than larger particles (Lai et al., 2007, Norris et al., 1998). Variability of PDI was very high both in 4 and 6 minutes of sonication time, creating a problem for the reproducibility in industrial production of nanoparticles. In addition of that high variability causes low absorption and poor targeted delivery of lycopene to the organ. Mohsin et al. (2019) inferred that increased sonication time leads to a decreased size of the polymeric particles. Sari et al. (2015) identified that surfactants significantly contributed to the encapsulation efficiency of bioactive compounds in polymeric nanoparticles.

Increasing surfactant concentration had an inverse effect on the hydrodynamic diameter of the nanoparticles. It means that increasing surfactant concentration in the composite caused reduction of hydrodynamic diameter of the nanoparticles which further help high absorption of lycopene nanoparticles based on Lai et al. (2007) and Norris et al., 1998 observation. A mixed effect appeared on other physicochemical properties of nanoparticles due to the surfactant concentration changes. Surfactant concentration did not affect PDI where increasing surfactant concentration from 300 mg to 400 mg improved the nanoparticles' stability (zeta potential), but further increment to 500 mg caused erosion significantly which further could decrease the control release activity of lycopene into the blood stream. Encapsulation efficiency never changed due to the increment of surfactant concentration from 300 to 400 mg, but a significant reduction appeared when 500 mg of surfactant was used to formulate nanoparticles. So, 300 mg of surfactant was the optimized condition for getting the highest encapsulation efficiency which facilitating the highest absorption profile for lycopene. Chuacharoen and Sabliov (2016) observed a narrow PDI of 0.18 with the lowest surfactant concentration for formulating the curcumin nanoemulsions. Chuacharoen et al. (2019) also observed an increased PDI (from 0.19 to 0.23) when a factor of 10 and 30 increased surfactant concentration. So lowest surfactant concentration should be selected to identify the optimum conditions as high surfactant could be health hazardous in the long-time consumption.

Increasing the polymer concentration also increases the hydrodynamic diameter and conductivity of the nanoparticles, which are undesirable for better bioaccessibility of the nanoparticles. Moreover, a lower hydrodynamic diameter has higher absorption and higher conductivity, facilitating a higher degradation effect against the latest heat treatments like microwave, ohmic, and pulse electric fields. However, PDI did not change polymer concentration. No effect appeared on the stability and encapsulation efficiency of the nanoparticles when polymer concentration had changed from 300 to 400 mg. However, 500 mg concentration had a deterioration and improvement effect on the stability and encapsulation efficiency of the encapsulated nanoparticles, respectively. So, polymeric concentration should not be high enough for high efficiency as it will compromise the stability of nanoparticles based on the homogeneity test results. On the other hand , high doses of polymer can increase the polymer load to the blood stream and further increase probable toxicity in the long run.

Multivariate regression analysis (Table-3) explained that surfactant concentration had an inverse relation (-23.1%) with the hydrodynamic diameter of the nanoparticles (P<0.05) where sonication time led the change of zeta potential (P<0.05) and encapsulation efficiency (P<0.05) by 63.8% and 85.7%, respectively. The conductivity of the nanoparticles is majorly controlled (44.9%) by changes in polymer concentration in the composite formula. So, it can be inferred that sonication time strongly controlled the nanoparticles' zeta potential and encapsulation efficiency, where surfactant and polymer concentration strongly controlled hydrodynamic diameter and conductivity of the encapsulated lycopene nanoparticles, respectively.

Table-3: Multiple Regression analysis for identifying the optimized composite

| Response Variables | Factors | Standardized Coefficients Beta | t | Sig. | Correlations Zero-order | Partial | Part |
|---|---|---|---|---|---|---|---|
| Hydrodynamic Diameter | Surfactant | -0.231 | -2.121 | 0.037 | -0.231 | -0.235 | -0.231 |
| Zeta Potential | Sonication Time | 0.638 | 7.394 | 0.000 | 0.638 | 0.644 | 0.638 |
| Conductivity | Sonication Time | 0.250 | 2.564 | 0.012 | 0.250 | 0.281 | 0.250 |
|  | Polymer Conc. | 0.449 | 4.610 | 0.000 | 0.449 | 0.465 | 0.449 |
| Encapsulation | Sonication Time | 0.857 | 16.419 | 0.000 | 0.857 | 0.882 | 0.857 |

|  | Efficiency | Surfactant Conc. | -0.203 | -3.888 | 0.000 | -0.203 | -0.405 | -0.203 |

### 3.3. Optimization of Hydrodynamic Diameter:

Figure-3 identified that the lowest hydrodynamic observed at 6 minutes of sonication time, 500 mg of surfactant, and 300 mg of polymer concentration. Lai et al. (2007) observed that nanoparticles sized between 100 and 200 nm had a moderate absorption through the intestinal lining. Based on the literature and homogeneity test results of the present study, 5 minutes of sonication time, 300 mg of surfactant, and 300 mg of polymer concentration could be an optimum condition for getting the 100 to 200 nm of nanoparticles. On the other hand, all other composites need higher surfactant, polymer, or sonication time for getting the same hydrodynamic profile (100-200 nm) of the nanoparticles.

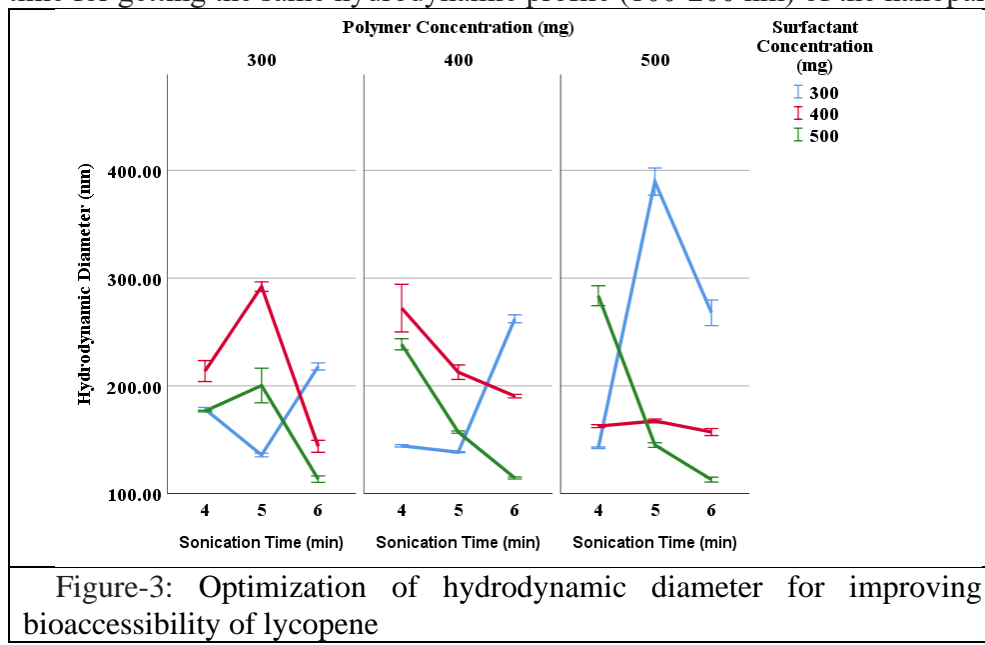

Figure-3: Optimization of hydrodynamic diameter for improving bioaccessibility of lycopene

Many studies worked with high surfactant concentration (Joung et al., 2016; Chuacharoen and Sabliov, 2016) compared to concentration generally used for nanoparticle synthesis (Chuacharoen and Sabliov, 2017) which had detrimental effect on the delivery of the entrapped bioactive (Barzegar, A.; Moosavi-Movahedi, 2011). Excessive surfactant also induces undesirable characteristics in foods fortified with these components (Barzegar and Moosavi-Movahedi, 2011). For example, excessive surfactant caused undesirable off-flavor lipid oxidation of milk during the processing and storage of fabricated nano-emulsion systems (Haidar et al., 2017). Chuacharoen et al. (2019) observed that the lowest surfactant contributed the largest size of nanoparticles for the curcumin. Chuacharoen and Sabliov (2016) found that at lowest surfactant concentration, tested for curcumin nanoemulsions, had the largest size of 193.93 nm.

### 3.4. Optimization of Polydispersity Index:

The polydispersity index (PDI) is essential for identifying the probability of nanoparticles absorption through the intestine lining. The lower the PDI, the higher the absorption through the intestinal lumen, and the higher the probability of getting highly efficient targeted delivery of encapsulated bioactive compounds. The lowest PDI value was found when 6 minutes of sonication time was used, incorporating with 400 mg and 500 mg of surfactant and polymer concentration, respectively.

Moreover, a very low PDI (slightly above 0.2) value was also found with lower surfactant concentration (300 mg), 5 minutes of sonication time, and 300 mg of polymer concentration which is more desirable than higher surfactant and polymer concentration. So, 300 mg of surfactant should be considered an optimized condition as hydrodynamic diameter of that particular composite was also suitable for moderate absorption. Haidar et al. (2017) observed that the lower the surfactant concentration higher the stability of fortified media used for delivering nutrients through diet.

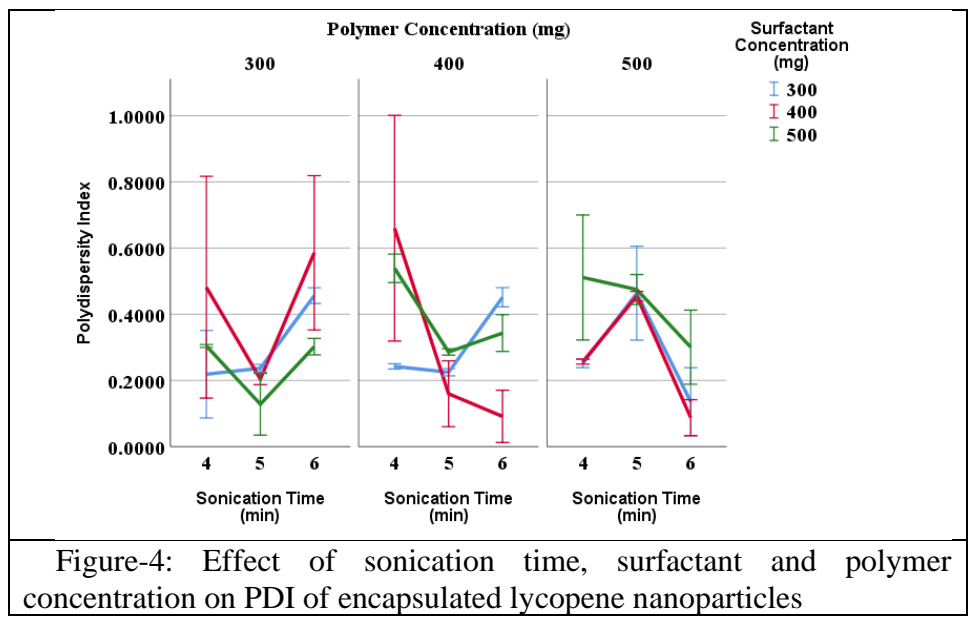

Figure-4: Effect of sonication time, surfactant and polymer concentration on PDI of encapsulated lycopene nanoparticles

### 3.5. Optimization of zeta potential value

According to Wissinga et al. (2004) and Jacobs et al. (2000) zeta potential values of greater than 30 mV of charge provide good stability, whereas 60 mV gives excellent stability. Faster aggregation can be termed as ZP of -5 mV to +5 mV, but greater than 20 mV will provide only short-term stability.

Higher the zeta potential explains the higher stability of the nanoparticles. According to Figure-5, all the compositions had a charge greater than 60 mV, where the highest ZP was found at 4 minutes of sonication time, 400 mg of surfactant, and 400 mg of polymer concentration. Chuacharoen and Sabliov (2016) observed that zeta potential values were less negative (-54.27 and -48 mV) at increasing lecithin because it induced a stronger anionic reaction resulting in a decreased negative surface charge. As 5 minutes of sonication time, 300 mg of surfactant and 300 mg polymer concentration provided better hydrodynamic diameter and PDI value incorporating with a charge of greater than 60 mV so, that should be the optimum condition considering the three physicochemical (Hydrodynamic diameter, PDI and zeta potential) properties of formulated nanoparticles. Moreover, Honary and Zahir (2013) observed that nanoparticles with slight negative change and 150 nm in diameter more efficiently accumulate in tumor cells than positive change or more significant in diameter which also supporting the 5 minutes of sonication time, 300 mg of surfactant and 300 mg of polymer concentration to be the optimized one. Additionally, positive charged NPs appear to have greater phagocytic uptake than negatively charged and that helps to excrete positive charged particles from the blood stream through urine and increase the targeted delivery of negatively charged particles component (Honary and Zahir, 2013).

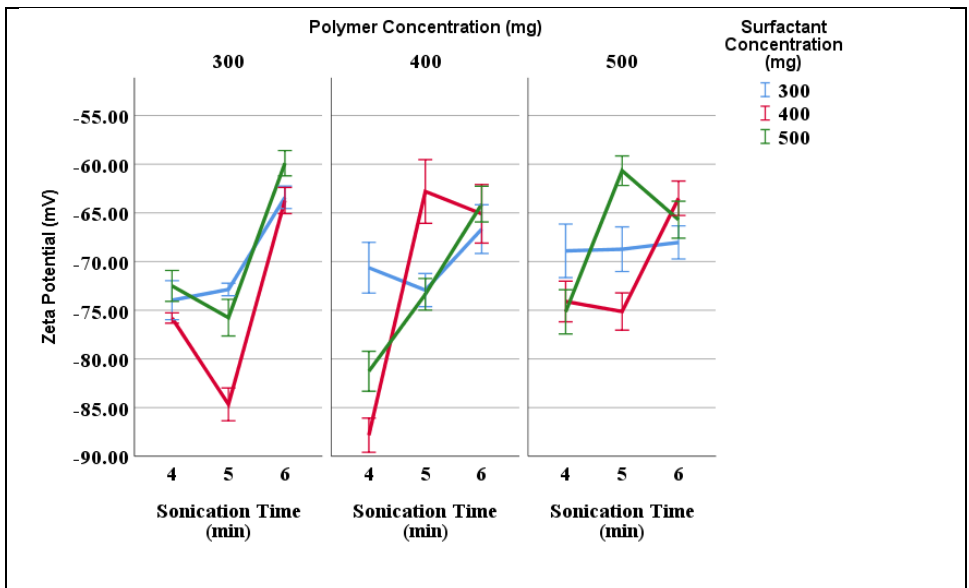

Figure-5: Effect of sonication time, surfactant, and polymer concentration on zeta potential of lycopene-PLGA nanoparticles

### 3.6. Optimization of Conductivity of Lycopene Nanoparticles:

Higher the conductivity facilitates the damage due to heat treatment such as microwave, pulse electric field, or ohmic heating, as these treatments depend on the food materials' electron flow behavior. From figure-6 the lowest electric conductivity was 5 minutes of sonication time, 400 mg of polymer, and 300 mg of surfactant concentration. The electric conductivity of the blend increases with an increase in sonication time until a certain point (Mohsin et al., 2019). So, 5 minutes of sonication time and 300 mf of surfactant concentration should be the optimized condition for best conductivity of the lycopene nanoparticles.

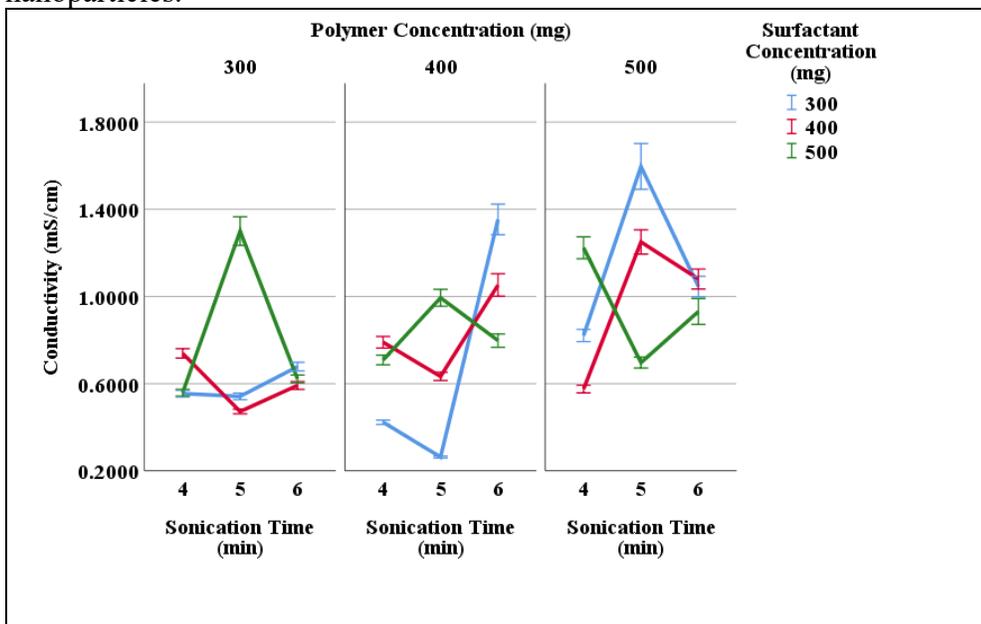

Figure-6: Effect of sonication time, surfactant, and polymer concentration on conductivity of lycopene-PLGA nanoparticles

### 3.7. Optimization of Encapsulation Efficiency Lycopene Nanoparticles

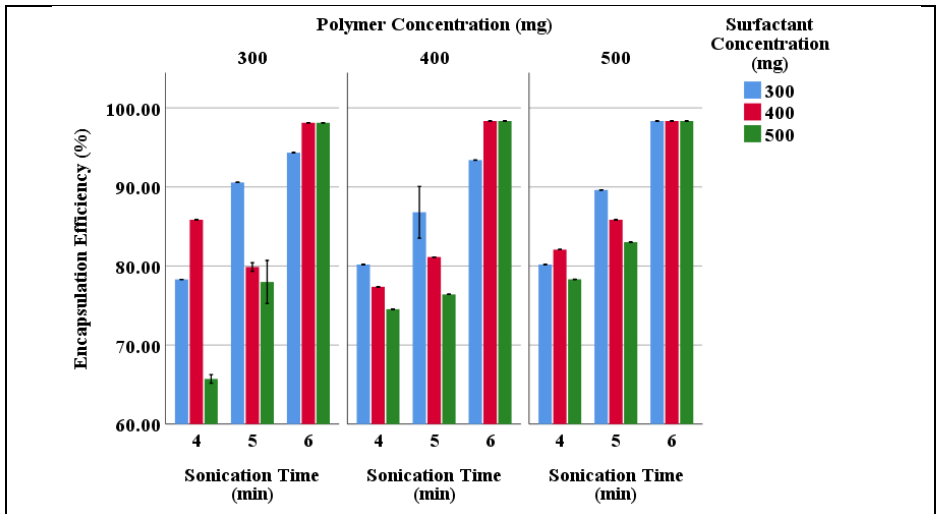

Figure-7: Effect of surfactant, polymer concentration and sonication time on encapsulation efficiency of lycopene in polymeric nanoparticles

Figure-7 depicted the surfactant, polymer concentration, and sonication time effect on the encapsulation efficiency of lycopene. ANOVA test results showed (Figure-7) that the interaction effect of surfactant, polymer concentration and sonication time significantly affects lycopene's encapsulation efficiency in polymeric nanoparticles. Figure-7 also explained the highest efficiency found at 6 minutes of sonication time irrespective of polymer or surfactant concentrations. Wu et al. (2017) observed that the addition of surfactant could improve the encapsulation efficiency of curcumin into the polymer, but which could be health hazardous in the long-time use. Therefore, it can be concluded that the droplet size, distribution, surface charge, and encapsulation efficiency of bioactive compounds in polymeric solution depended on the ratio of surfactant to the total composite of that particular system.

Sonication time of 5 minutes together with 300 mg of surfactant and 300 mg of polymer concentration contributed more than 90 percent of encapsulation efficiency of lycopene in polymeric nanoparticles without having low conductivity, excellent stability (zeta potential value of more than 60 mV) and low PDI (close to 0.2) must be the optimum condition considering all physicochemical properties of encapsulated nanoparticles required to the highest absorption.

### 3.8. Physical and Morphological Characteristics of Optimized Nanoparticles:
### 3.8.1. Raman Spectroscopy Analysis for Optimized Nanoparticles:

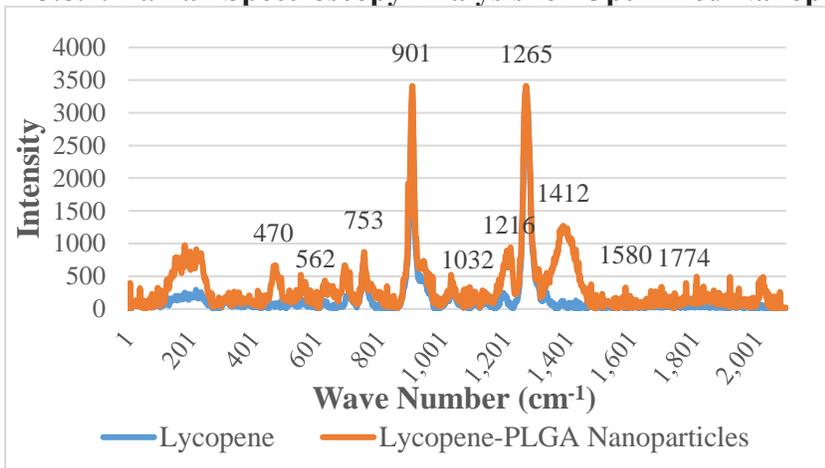

Figure 8: Raman spectroscopy analysis of standard and encapsulated lycopene nanoparticles

Figure 8 is the confirmatory analysis for encapsulation of lycopene into the optimized nanoparticles. The band at 1412 $cm^{-1}$ expresses the intensity peak for glycolic acid, which appears on the nanoparticles as PLGA has a monomer of glycolic acid. Low-intensity peak found at 1774 $cm^{-1}$ which is the identification characteristics of the ester bond in PLGA build between lactic acid and glycolic acid (van Apeldoorn et al., 2004). However, intensity peaks at 1032 cm$^{-1}$ and 1580 cm$^{-1}$ are some identifying characteristics of lycopene visible on both lycopene and lycopene-PLGA nanoparticle's intensity curve (Radu et al., 2016; Qin et al., 2012; Nikbakht et al., 2011) which validate that lycopene was present into the final encapsulated nanoparticles.

### 3.8.2. Physical and Morphological Characteristics of Optimized Nanoparticles:

Figure-9 (a) explained that hydrodynamic diameter distribution of optimized nanoparticles was normally distributed which is a potential characteristic for target delivery of encapsulated bioactive compounds. A sharp peak appeared at the middle of the distribution, which means uniformity among different particle's diameters. Normal distribution with the highest peak at the middle of the curve also appeared for the zeta potential value distribution of optimized nanoparticles (Figure-7 (b)). SEM (Scanning Electron Microscopy) (Figure-7 (c), (d)) imaging appeared that encapsulated lycopene nanoparticles were spherical which is highly favorable for intestinal absorption through the epithelial cells of intestinal lining.

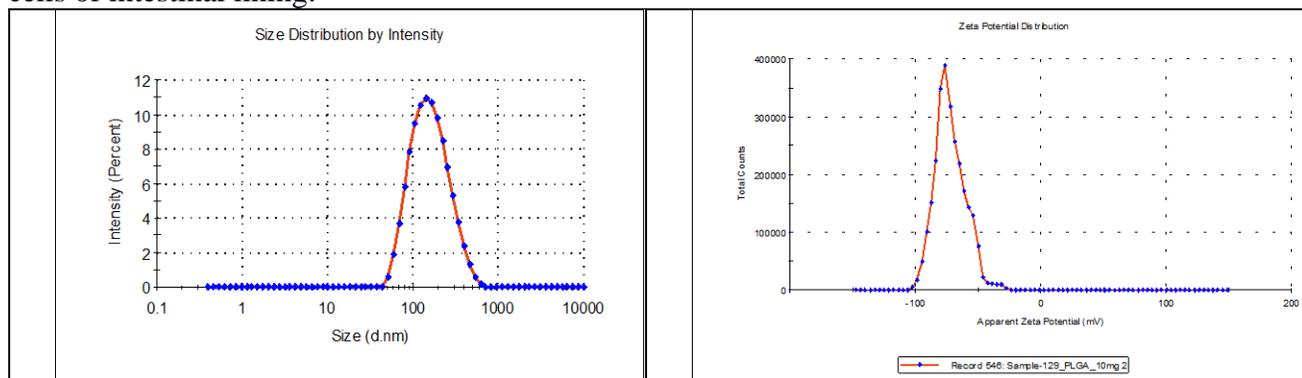

Figure-9 (a): Hydrodynamic Diameter Distribution of the optimized lycopene-PLGA nanoparticles

Figure-9 (b): Zeta Potential Distribution of the optimized lycopene-PLGA nanoparticles

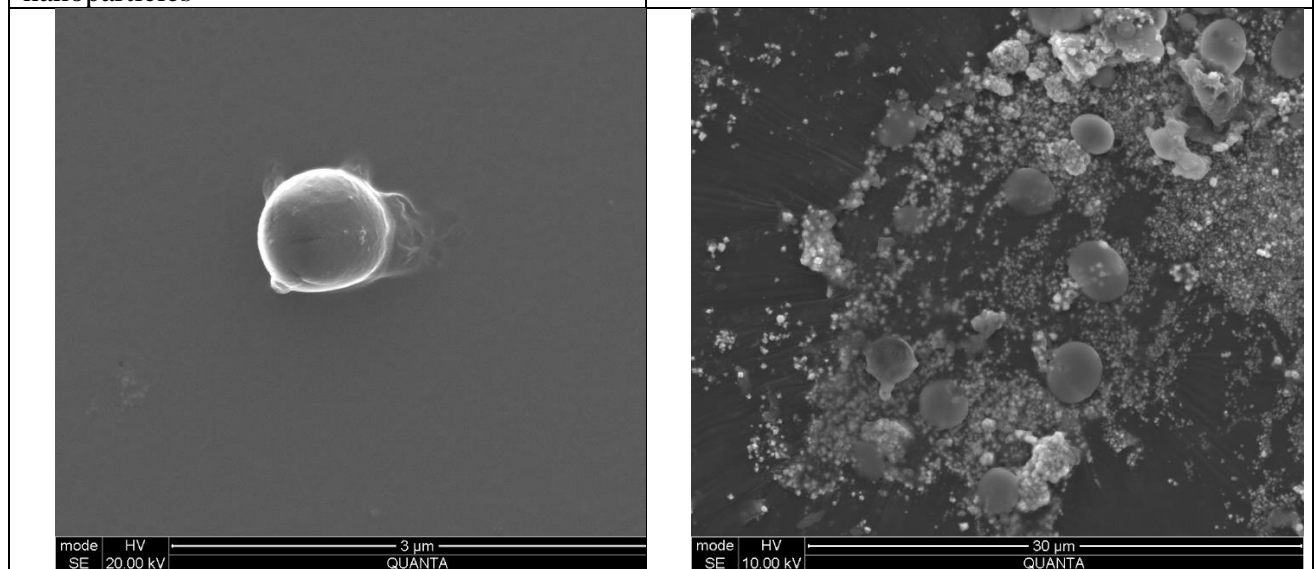

Figure-9 (c): SEM Imaging of Optimized Lycopene-PLGA nanoparticles

### 3.9. In-vitro Control Release Kinetics of the Optimized Lycopene-PLGA-NPs:

PLGA-Lycopene NPs released lycopene (Figure-10 (a)) at a reasonably constant rate until the first 12 days of the experiment. During the 12 days, it released 53.89 ± 6.65 % of the total content (at a relatively constant rate of 4.49 percent/day) into the invitro phosphate buffer solution (PBS, pH =7.4) system at 37°C. Burst release was observed on the 13th day, where 41.7 ± 3.75 % of the total content was released within a single day. The biodegradable polymers usually follow three distinct mechanisms for drug release, i.e., desorption of drug from NP's surface, diffusion from the polymer matrix, and re-adsorption, degradation, and dissolution/erosion of the polymeric network. Ahmed et al. (2020) observed a burst release of bioactive compounds through PLGA initially, whereas Khan et al. (2018) found consistent release when chitosan was used to make an additional barrier for PLGA NPs. Shadab et al. (2014) found similar results in PLGA nanoparticles where burst release occurred within 30 min of experimental condition, and sustained release appeared over the next 25 days of experimental time. According to Korsmeyer - Peppas Model, the release kinetics of the PLGA-lycopene nanoparticles for the present study follow the following equation with 96% of accuracy (Figure-10 (b)):

$Y = 3.81X^n$, ($R^2 = 0.96$)   -------------------------------(5)

Where, Y = Fraction of lycopene released
X = Time in days
n = the release exponent indicative of the mechanism of transport of drug through the polymer.

According to Korsmeyer-Peppas mathematical model fitness test (Figure-10 (b)), an exponential value of n= 0.99 explained that the release kinetics followed the no-Fickian type of diffusion. No-Fickian types of diffusion are diffusion where the release of bioactive compounds happens due to erosion of polymer from the surface of the encapsulated polymeric nanoparticles (Paarakh et al., 2018). The $R^2$ value was 0.96, which means the control release kinetics of PLGA-lycopene satisfied 96% of confidence of fitness to the Korsmeyer-Peppas mathematical model.

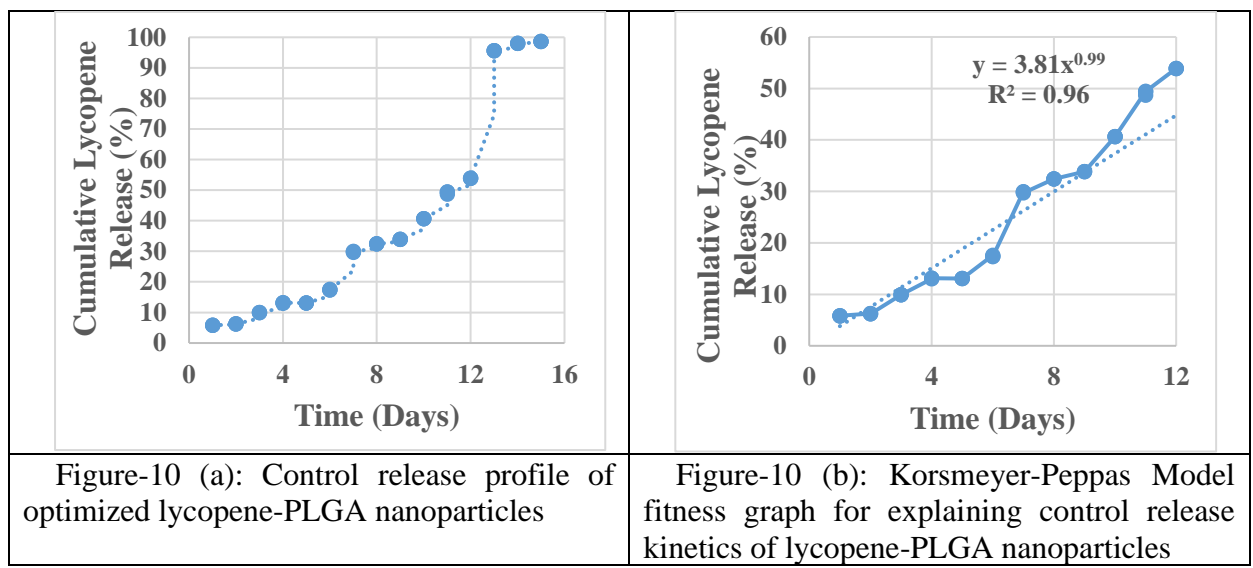

| Figure-10 (a): Control release profile of optimized lycopene-PLGA nanoparticles | Figure-10 (b): Korsmeyer-Peppas Model fitness graph for explaining control release kinetics of lycopene-PLGA nanoparticles |

## 3.10. Bioaccessibility of Lycopene

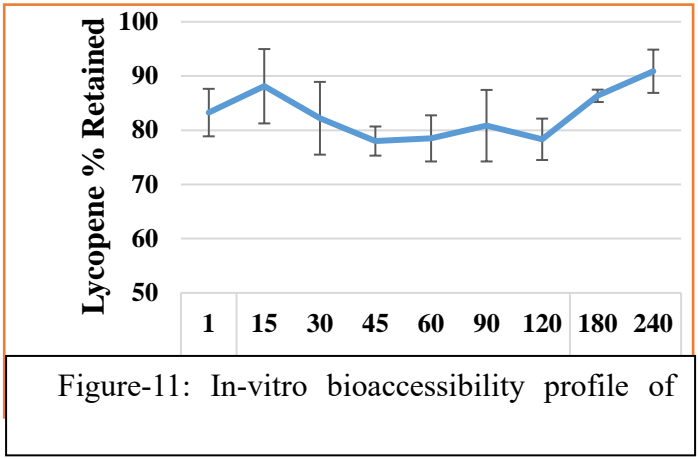

Figure-11: In-vitro bioaccessibility profile of

### 3.10.1 In-vitro digestion of lycopene nanoparticles:

The bioaccessibility study was done based on the DPPH activity retention capacity of lycopene throughout the simulated in-vitro human digestion. After completion of in-vitro digestion of optimized lycopene nanoparticles, no significant change had observed in the DPPH activity of the lycopene from mouth to intestine. It means that PLGA used to encapsulate the lycopene effectively protects the autooxidation or isomerization reaction, observed by Petyaev (2016) and Tommonaro et al. (2017), of lycopene due to digestive enzymes. The bioaccessibility of lycopene was 90.87±3.99% (Figure-9) which was 3-9 times greater than that of non-capsulated lycopene whose absorption was between 10 and 30% (Rao & Agarwal 1999; Gärtner, 1997).

### 3.10.2. Absorption profile of optimized lycopene nanoparticles:

Figure-10 shows the absorption behavior of lycopene after invitro digestion of the nanoparticles. After the digestion 80% of the total nanoparticles usually absorbed through Duodenum by direct diffusion. Based on the Korsmeyer-Peppas fitness test results it can be inferred that the absorption profile of PLGA lycopene nanoparticles was followed the below mentioned equation.

$M_t/M_\alpha = kt^n$

$y = 29.63t^{0.7116}$ (Figure-12)

$M_t/M_\alpha$: fraction of drug released at time $t$ = y
$k$: Korsmeyer-Peppas constant = 29.63
$t$: time
$n$: release exponent = 0.7116

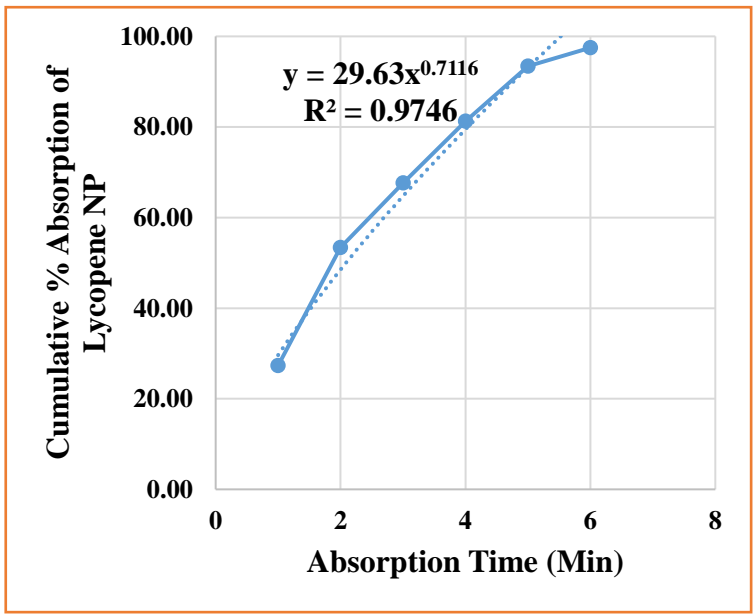

The correlation coefficient ($R^2$) = 97.46% indicated that the absorption behavior of lycopene nanoparticles followed the Korsmeyer-Peppas model by 97.46%. The release exponent =0.7116, which means that the lycopene PLGA NPs followed non-Fickian types of diffusion (Ravi and Mandal, 2015).

### CONCLUSIONS

Encapsulation has become a prominent technique to improve the bioaccessibility of bioactive compounds like lycopene. Moreover, PLGA can be a potential carrier to improve the bioaccessibility of lycopene through diet. Research evident that physicochemical properties of the encapsulated nanoparticles directly affect the bioaccessibility or bioavailabilty of the lycopene. Researchers proved that sonication time, surfactant, and polymer concentration were potential contributing factors for formulating encapsulated lycopene nanoparticles. However, the sonication time strongly controls the stability (zeta potential value) and encapsulation efficiency (63.8% and 85.7%, respectively) of the lycopene nanoparticles where surfactant (-20.3%) and polymer (49.9%) concentration strongly (P<0.05) control the hydrodynamic diameter and conductivity of the nanoparticles. After optimizing the sonication time, surfactant, and polymer concentration, bioaccessibility of lycopene due to encapsulation was observed to be 90.87±3.99% which is 3-9 times greater than that of normal lycopene ingested through the oral route. Furthermore, both control release kinetics (R=96%) and absorption (R=97.46%) profiles of the lycopene nanoparticles followed non-Fickian types of diffusion. Hence, food industries can use the lycopene nanoparticles to produce new functional foods to provide high bioaccessible and prolonged circulatory lycopene through diet to prevent chronic diseases in future.

**ACKNOWLEDGEMENTS**

USDA Evan-Allen Accession No. 1013057: Project Title: Modeling in Vitro Control Release and Diffusion of Loaded Nanoparticles (LNP) in the GI tract - Impact of processing.